\documentclass[sigconf]{acmart}
\AtBeginDocument{%
  }

\usepackage{xcolor}
\usepackage{pifont}
\usepackage{enumitem}
\usepackage{hyperref}
\usepackage{algorithm}
\usepackage{algpseudocode}
\usepackage{graphicx}
\usepackage{array} 
\usepackage{multirow}
\usepackage[table]{xcolor}
\usepackage{colortbl}
\usepackage{threeparttable}
\usepackage[most]{tcolorbox}
\usepackage{listings}
\usepackage{lipsum}
\tcbuselibrary{listingsutf8}
\usepackage{nameref}
\usepackage[commandnameprefix=ifneeded]{changes}
\tcbuselibrary{listings,breakable}
\newcounter{myboxctr}
\newtcolorbox{mybox}[3][]{
  floatplacement={#3},
  float,
  enhanced,
  colback=white, colframe=black!50,
  left=8pt,
  top=2pt,
  right=7pt,
  bottom=1pt,
  after skip=2pt,
  fonttitle=\small\bfseries,
  fontupper=\scriptsize,
  breakable,
  before title={\refstepcounter{myboxctr}\label{#1}},
  title=Box~\themyboxctr: #2,
}

\usepackage[framemethod=tikz]{mdframed}
\mdfdefinestyle{customstyle}{
  linecolor=gray!100,
  linewidth=3pt,
  innerleftmargin=3pt,
  topline=false,
  rightline=false,
  bottomline=false,
  leftline=true,
  innerrightmargin=3pt,
  innertopmargin=3pt,
  innerbottommargin=3pt,
  backgroundcolor=gray!15
}
\newenvironment{custommdframed}
  {\begin{mdframed}[style=customstyle]}
  {\end{mdframed}}

\newcommand{\cmark}{\textcolor{green!60!black}{\ding{51}}} 
\newcommand{\xmark}{\textcolor{red}{\ding{55}}} 
\definecolor{DeepGreen}{RGB}{0,128,0}
\definecolor{DeepRed}{RGB}{178,34,34}
\definecolor{lightgray}{gray}{0.92}

\definecolor{codegreen}{rgb}{0,0.6,0}
\definecolor{codegray}{rgb}{0.5,0.5,0.5}
\definecolor{codepurple}{rgb}{0.58,0,0.82}
\definecolor{backcolour}{rgb}{0.95,0.95,0.96}
\definecolor{framecolor}{rgb}{0.8,0.8,0.8}

\lstdefinestyle{mystyle}{
    basicstyle=\ttfamily\tiny,    
    backgroundcolor=\color{backcolour},   
    frame=single,                         
    frameround=tttt,                      
    rulecolor=\color{framecolor},         
    numbers=left,                         
    numberstyle=\tiny\color{codegray},    
    stepnumber=1,                         
    numbersep=8pt,                        
    commentstyle=\color{codegreen},       
    keywordstyle=\color{magenta},         
    stringstyle=\color{codepurple},       
    identifierstyle=\color{blue},         
    emphstyle=\color{orange},             
    tabsize=4,                            
    showspaces=false,                     
    showstringspaces=false,               
    breaklines=true,                      
    breakatwhitespace=true,               
    captionpos=b,                         
    xleftmargin=5pt,                      
    xrightmargin=2pt,                     
    aboveskip=5pt,                        
    belowskip=5pt                         
}
\lstdefinestyle{Python}{
    language=Python,
    morekeywords={as, async, await, False, None, True, pass, self, lambda},
    deletekeywords={print},  
    keywordstyle=\color{blue}\bfseries,
    commentstyle=\color{codegreen}\itshape,
    stringstyle=\color{codepurple},
    emph={__init__,__name__,__main__,__file__}, 
    emphstyle=\color{red}\bfseries
}

\newcommand{\ourmodel}{\textsc{RepoScope}}

\setcopyright{acmlicensed}
\copyrightyear{2018}
\acmYear{2018}
\acmDOI{XXXXXXX.XXXXXXX}
\acmConference[Conference acronym 'XX]{Make sure to enter the correct
  conference title from your rights confirmation email}{June 03--05,
  2018}{Woodstock, NY}
\acmISBN{978-1-4503-XXXX-X/2018/06}

\begin{document}

\title{\ourmodel{}: Leveraging Call Chain-Aware Multi-View Context for Repository-Level Code Generation}


\author{Yang Liu$^{1}$, Li Zhang$^{1}$, Fang Liu$^1$$^\ast$, Zhuohang Wang$^1$, Donglin Wei$^1$, Zhishuo Yang$^1$ \\ Kechi Zhang$^2$, Jia Li$^2$, Lin Shi$^3$}
\thanks{$^{\ast}$Corresponding author.}
\affiliation{%
\institution{$^1$State Key Laboratory of Complex \& Critical Software Environment, School of Computer Science and Engineering, Beihang University, Beijing, China \\ $^2$School of Computer Science, Peking University, Beijing, China \\ $^3$School of Software, Beihang University, Beijing, China}
  \country{}
}

\email{{liuyang26, fangliu}@buaa.edu.cn}

\renewcommand{\shortauthors}{Liu et al.}

\begin{abstract}
Repository-level code generation aims to generate code within the context of a specified repository. Existing approaches typically employ retrieval-augmented generation (RAG) techniques to provide LLMs with relevant contextual information extracted from the repository. However, these approaches often struggle with effectively identifying truly relevant contexts that capture the rich semantics of the repository, and their contextual perspectives remains narrow. 
Moreover, most approaches fail to account for the structural relationships in the retrieved code during prompt construction, hindering the LLM's ability to accurately interpret the context.
To address these issues, we propose \ourmodel{}, which leverages call chain-aware multi-view context for repository-level code generation. \ourmodel{} constructs a Repository Structural Semantic Graph (RSSG) and retrieves a comprehensive four-view context, integrating both structural and similarity-based contexts. We propose a novel call chain prediction method that utilizes the repository's structural semantics to improve the identification of callees in the target function. Additionally, we present a structure-preserving serialization algorithm for prompt construction, ensuring the coherence of the context for the LLM. Notably, \ourmodel{} relies solely on static analysis, eliminating the need for additional training or multiple LLM queries, thus ensuring both efficiency and generalizability. Evaluation on widely-used repository-level code generation benchmarks (CoderEval and DevEval) demonstrates that \ourmodel{} outperforms state-of-the-art methods, achieving up to a 36.35\% relative improvement in pass@1 scores. Further experiments emphasize \ourmodel{}'s potential to improve code generation across different tasks and its ability to integrate effectively with existing approaches. We provide the replication package at \url{https://github.com/Lorien1128/RepoScope}.
\end{abstract}



\copyrightyear{2026}
\acmYear{2026}
\setcopyright{cc}
\setcctype{by}
\acmConference[ICSE '26]{2026 IEEE/ACM 48th International Conference on Software Engineering}{April 12--18, 2026}{Rio de Janeiro, Brazil}
\acmBooktitle{2026 IEEE/ACM 48th International Conference on Software Engineering (ICSE '26), April 12--18, 2026, Rio de Janeiro, Brazil}
\acmPrice{}
\acmDOI{10.1145/3744916.3773211}
\acmISBN{979-8-4007-2025-3/26/04}


\maketitle

\section{Introduction}
Code generation can substantially reduce manual effort by automating repetitive and time-consuming programming tasks, offering transformative potential for software engineering.
In recent years, Large Language Models (LLMs), such as DeepSeek-Coder \cite{guo2024deepseekcoder}, QwenCoder \cite{qwencoder}, GPT-4o \cite{achiam2023gpt4}, \textit{etc.}, have demonstrated impressive capabilities in code generation tasks. 
However, their performance often degrades significantly when applied to repository-level code generation tasks within complex, large-scale repositories.
This challenge arises primarily from two factors: \ding{182} \textbf{constrained context windows} limiting the model's access to the broader project context needed to understand architecture and intricate dependencies, and \ding{183} \textbf{a lack of repository-specific knowledge} \cite{shrivastava2023repository}, including details of internal APIs and domain-specific coding conventions not captured within the immediate context window.

To mitigate the above challenge, retrieval-augmented generation (RAG) \cite{gao2023retrieval} approaches have been introduced, aiming to provide LLMs with relevant contextual information extracted from the repository.
Early approaches \cite{zhang2023repocoder, shapkin2023dynamic} retrieve textually similar code snippets from the repository using the code context as a query.  
Recent methods have sought to refine retrieval through structural and semantic augmentation, \textit{e.g.}, by incorporating imported definitions \cite{liang2024repofuse}, performing dataflow analysis \cite{cheng2024dataflow}, or constructing repository‐level semantic graphs with GNNs for relevance scoring \cite{phan2024repohyper}. Agent-based pipelines further extend these capabilities by enabling iterative retrieval and LLM-guided evaluation of retrieval targets \cite{bi2024iterative,zhang2024codeagent,ma2024alibaba}. Such advances have yielded notable performance gains in repository-level code generation tasks. 
Nevertheless, existing approaches still exhibit critical limitations:

\noindent \textbf{Insufficient Relevance of Retrieved Context}. In RAG frameworks, the quality of LLM-generated code hinges critically on identifying the highly relevant context. While similarity-based retrieval approaches \cite{zhang2023repocoder, di2403repoformer, gao2024preference, wang2024rlcoder, liu2024graphcoder} can retrieve highly similar code, they overlook the structural semantics embedded in the repository.
This often leads to the omission of critical context, potentially causing generated code to conflict with repository-specific knowledge, such as the usage of constants or APIs.
To mitigate this issue, structural augmentation approaches are proposed \cite{liang2024repofuse,cheng2024dataflow,bi2024iterative}. However, existing methods typically rely on shallow structural semantics, such as basic dependency parsing or dataflow analysis. As a result, they struggle to effectively assess and filter the relevance of retrieved code, hindering the precise identification of truly relevant context.

\noindent \textbf{Limited contextual perspectives.} When encountering a new project, human programmers typically analyze related code from multiple perspectives to understand the behavior of a particular snippet. For instance, they may examine where it is invoked, how its return values are handled, which function it calls, and what structurally similar code exists. In contrast, existing approaches often rely on just one or two such perspectives, hindering accurately modeling of rich semantics within repositories.
The rapid advancement in LLM capabilities now enables and demands repository-level code generation to incorporate richer, more diverse contextual views.

Furthermore, existing approaches exhibit several additional limitations. In terms of efficiency, approaches that depend on extensive supervised training \cite{phan2024repohyper,wang2024rlcoder} or on agent-based pipelines with multiple LLM queries \cite{zhang2024codeagent,ma2024alibaba} incur substantial computational/temporal cost, and often struggle to adapt across diverse repositories.
Regarding prompt organization, existing approaches typically construct prompts by simply concatenate the retrieved code elements, either directly or in a predetermined order \cite{phan2024repohyper,liang2024repofuse, cheng2024dataflow}, overlooking the inherent structural relationships between them (\textit{e.g.}, class-method hierarchies). The lack of structural representation may hinder LLMs' ability to accurately interpreting context semantics, potentially degrading code generation quality.

\begin{table}[t]
\centering
\setlength{\abovecaptionskip}{0.1cm}
\caption{Comparison between \ourmodel{} and existing methods, including whether the method ranks retrieved code elements to reduce irrelevant context, utilizes more than two contextual views, operates without additional training, and performs code generation using only a single LLM query.}
\label{tab:method_compare}
\resizebox{\linewidth}{!}{
\begin{tabular}{p{0.25\linewidth}p{0.18\linewidth}p{0.15\linewidth}p{0.18\linewidth}p{0.20\linewidth}}
\toprule 
\textbf{Method} & \textbf{Element Ranking} & \textbf{Diverse Views} & \textbf{Training-Free} & \textbf{Single LLM Query} \\
\midrule
RepoCoder\cite{zhang2023repocoder} & \xmark & \xmark & \cmark & \xmark \\
RLCoder\cite{wang2024rlcoder} & \xmark & \xmark & \xmark & \cmark \\
RepoFuse\cite{liang2024repofuse} & \xmark & \xmark & \cmark & \cmark \\
DRACO\cite{cheng2024dataflow} & \xmark & \xmark & \cmark & \cmark \\
RepoHYPER\cite{phan2024repohyper} & \cmark & \xmark & \xmark & \cmark \\
CoCoGen\cite{bi2024iterative} & \xmark & \xmark & \cmark & \xmark \\
CodeAgent\cite{zhang2024codeagent} & \xmark & \cmark & \cmark & \xmark \\
LingmaAgent\cite{ma2024alibaba} & \cmark & \xmark & \cmark & \xmark \\
\rowcolor{lightgray} \textbf{\ourmodel{}} & \cmark & \cmark & \cmark & \cmark \\
\bottomrule
\end{tabular}
}
\vspace{-0.4cm}
\end{table}

To this end, we propose \ourmodel{}, a framework that leverages call chain–aware multi-view context for repository-level code generation. For each repository, we construct a Repository Structural Semantic Graph (RSSG).
With the RSSG, \ourmodel{} retrieves two types of structure-based context for the target function: its direct callers and its potential callees. 
These structure-based contexts are then combined with two additional similarity-based contexts. Together, they form a comprehensive four-view context, equipping the LLM with rich background knowledge needed for code generation.
Specifically, to identify potential callees, we propose a call chain prediction method, fully exploiting the structural semantics of the repository to predict entire call chains (\textit{i.e.}, a sequence of callees connected via semantic relations, as defined in Section \ref{sec:definitions}), rather than just isolated callees. This method enhances the accuracy of callee identification and thus provides context with stronger relevance.
When constructing the prompt with the retrieved four-view context, we introduce a structure-preserving serialization algorithm to convert the contexts into a coherent token sequence, preserving their original hierarchical relationships within the repository. This enables the LLM to accurately interpret context semantics, thus enhancing code generation quality.
Crucially, our approach relies solely on static analysis, requiring no additional training or multiple LLM queries, thereby ensuring both efficiency and generalizability. 
Table \ref{tab:method_compare} presents the comparison between \ourmodel{} and several representative repository-level code generation methods.

We evaluate \ourmodel{} on widely-used repo-level code generation benchmarks, CoderEval \cite{yu2024codereval} and DevEval \cite{li2024deveval}, using four advanced backbone LLMs and perform comprehensive comparisons with state-of-the-art baselines. The evaluation results demonstrate that \ourmodel{} outperforms the best-performing baseline across all backbone models, achieving up to 36.35\% relative improvement in pass@1 score. In addition, we further explored \ourmodel{}'s generalization capability and its integration with existing approaches. 
In summary, the contributions of this work are as follows: 
\begin{itemize}[leftmargin=*]
    \item We propose \ourmodel{}, a framework that utilizes call chain-aware multi-view context for repository-level function generation. By integrating context from four distinct perspectives, \ourmodel{} enriches the LLM with a more comprehensive understanding of the repository's semantics.
    \item We introduce Repository Structural Semantic Graph (RSSG), a heterogeneous directed graph constructed via static analysis, which captures various types of structural semantics within a code repository and enables more precise context retrieval.
    
    \item We introduce a call chain prediction approach which deeply leverages the structural semantics in the RSSG to predict the potential callees of the target function, thereby improving the relevance of the retrieved context. We further propose a structure-preserving serialization algorithm that converts context into coherent token sequences while preserving structural hierarchy.
    
    \item We conduct a comprehensive evaluation of \ourmodel{}, and the results demonstrate that it outperforms state-of-the-art methods, exhibits strong generalizability across different tasks, and offers potential for performance enhancement when integrated with existing approaches. 
\end{itemize}

\section{Preliminaries}

\begin{figure*}[t]
    \centering
    \setlength{\abovecaptionskip}{0.1cm}
    \includegraphics[width=\linewidth]{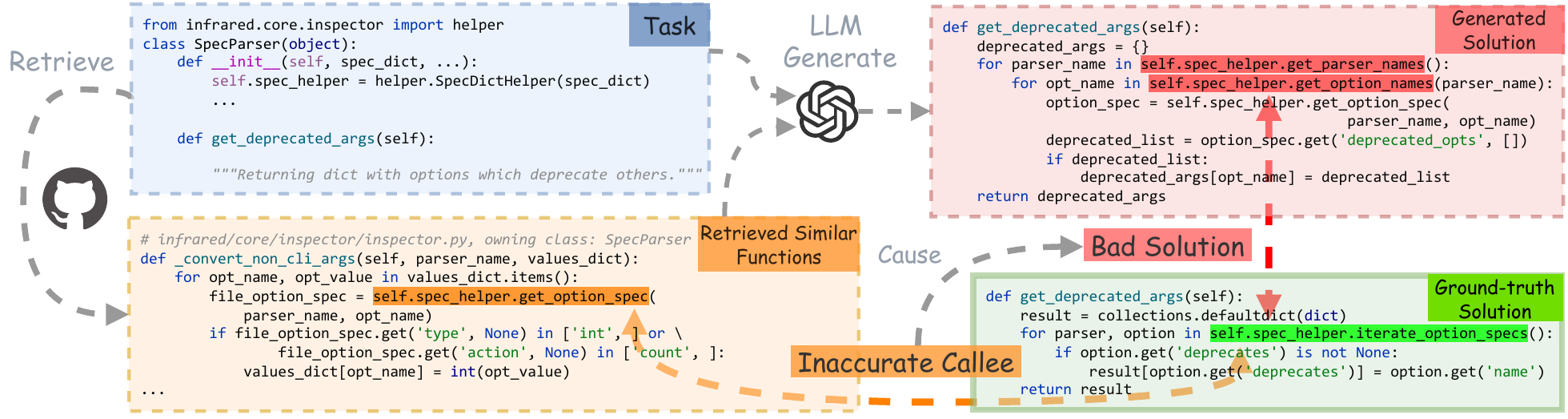}
    \caption{A motivating example. Similarity-based retrieval methods often struggle to accurately capture call-related information.}
    \label{fig:code_exp1}
    \vspace{-0.2cm}
\end{figure*}

\begin{figure}[h]
    \centering
    \setlength{\abovecaptionskip}{0.1cm}
    \includegraphics[width=0.8\linewidth]{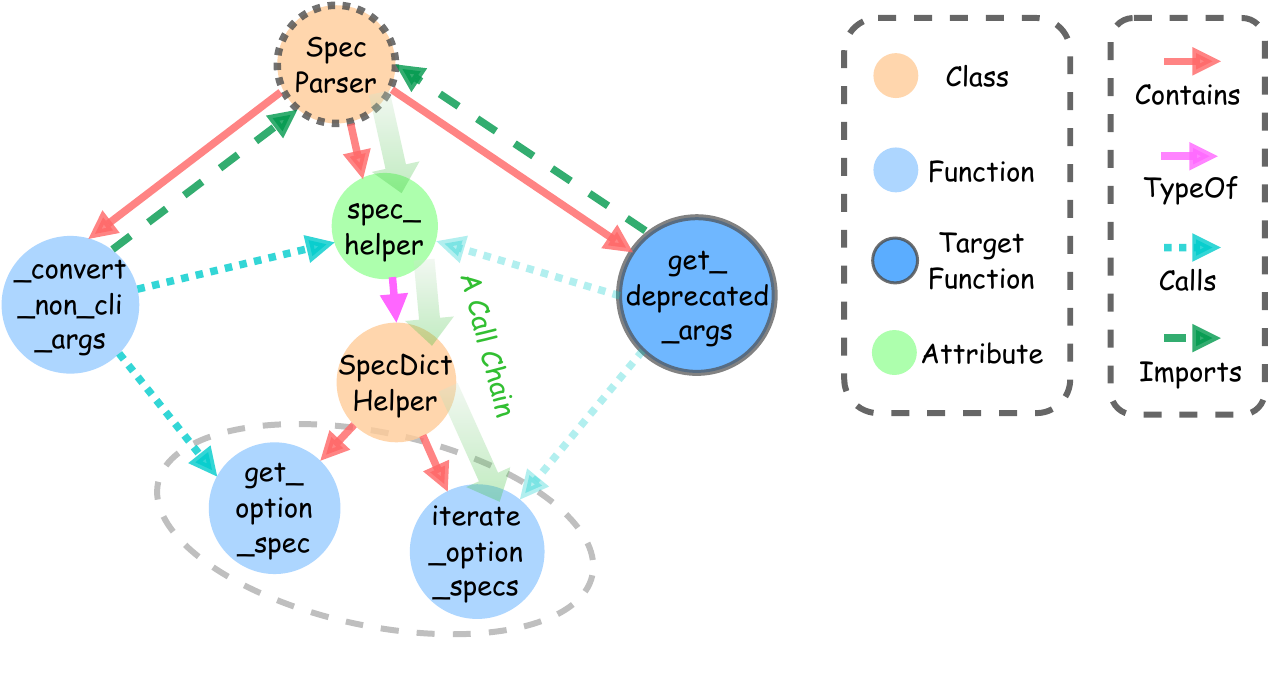}
    \caption{The RSSG derived from the code shown in Figure \ref{fig:code_exp1}.}
    \vspace{-0.3cm}
    \label{fig:graph_exp1}
\end{figure}

\subsection{Motivating Example}
\label{sec:motivating_example}

In programming languages, the relationships between various code elements (such as classes, methods/functions, and attributes) encode rich structural semantics that are vital for understanding code \cite{maletic2001supporting, plotkin2004origins, le1996structural}. These relationships include call relationships, structural and type dependency relationships, and import relationships, \textit{etc}. When implementing a new function, the most direct, precise and relevant information that can be obtained from these relationships typically includes the callers and callees of the target function \cite{zhao2015reusable, hassen2017scalable}. The callers reveal the function's intended use, providing insights into expected inputs and outputs, while the callees directly participate in constructing the function body itself. Additionally, type information for parameters and return values is crucial, which is often implicitly reflected in caller usage patterns. Critically, while static analysis can identify callers, callees present a distinct challenge.
\textit{Since the target function's body is unknown during generation, we cannot directly identify its callees from the repository. This necessitates \textbf{predicting} the likelihood of each code element being called by the target function}. The goal is to select the code elements with the highest probability to construct a highly relevant context for code generation. 
However, this task is non-trivial, requiring comprehensive analysis of type dependencies, code hierarchies, and cross-entity relations within the repository.

Existing RAG based code generation methods lack targeted design and thus often struggle to accurately retrieve or predict these code elements. As shown in Figure \ref{fig:code_exp1}, a similarity-based retrieval approach identifies the function \texttt{\_convert\_non\_cli\_args}, which is similar to the target function. While its call chain appears relevant, it erroneously terminates at \texttt{get\_option\_spec} instead of the required \texttt{iterate\_option\_specs}. When supplied with this inaccurate context, LLMs may hallucinate non-existent methods (\textit{e.g.}, \texttt{get\_parser\_names} and \texttt{get\_option\_names}), leading to incorrect solution. 
In contrast, proactively identifying the target function's dependency on \texttt{iterate\_option\_specs} and incorporating this information into the prompt would enhance generation accuracy.

More importantly, these code elements typically form sequences of interconnected calls, termed \textbf{Call Chain} (formally defined in Section \ref{sec:definitions}).
A call chain connects different code elements through type or structural relationships. As shown in Figure \ref{fig:code_exp1} and \ref{fig:graph_exp1}, a plausible call chain within the target function \texttt{get\_deprecated\_args} contains: \ding{172} The parent class \texttt{SpecParser} $\rightarrow$ \ding{173} The attribute \texttt{spec\_\\helper} $\rightarrow$ \ding{174} The attribute's type \texttt{SpecDictHelper} $\rightarrow$ \ding{175} The terminal function \texttt{iterate\_option\_specs}.
Compared to an isolated callee, a call chain encodes richer behavioral information and provides a more coherent contextual representation. Accurately predicting such chains is therefore highly beneficial for code generation.

It is also worth noting that, as shown in Figure \ref{fig:graph_exp1}, the correct callee (\texttt{iterate\_option\_specs}) and the incorrectly retrieved one (\texttt{get\_option\_spec}) also exhibit high similarity. This aligns with a common phenomenon in software development: \textit{similar functions within the same repository often exhibit similar calling behaviors}. Inspired by this insight, we propose a heuristic call chain prediction method that leverages the calling patterns of similar, existing functions to assist in predicting the call chain of the target function. This method will be detailed in Section \ref{sec:call_chain_prediction}.

\subsection{Definitions}
\label{sec:definitions}
This section formalizes key concepts foundational to our approach: \textit{Call}, \textit{Call Chain}, and our constructed \textit{Repository Structural Semantic Graph}.

\subsubsection{Call}
In this paper, \textit{Call} refers to the \textbf{access} or \textbf{invocation} of code elements within the repository that are accessible across scopes, including classes, functions/methods, attributes, \textit{etc}.

\subsubsection{Repository Structural Semantic Graph} 

The Repository Structural Semantic Graph (RSSG) is a multi-view heterogeneous directed graph that represents semantic relationships among code elements in the repository.
Specifically, RSSG includes three types of entities:
\begin{itemize}[leftmargin=*]
    \item \textbf{\textit{Class}} represents the classes defined in the code.
    \item \textbf{\textit{Function}} represents various functions, including both standalone functions and those bound to classes.
    \item \textbf{\textit{Attribute}} represents variables bound to classes.
\end{itemize}
Relations are categorized into three semantic views comprising six distinct types:

\begin{enumerate}[leftmargin=*]
    \item \textbf{\textit{Structural and Type Dependency}} relation defines the possible paths of a call chain. Entities involved in a sequence of directly related call operations can be linked through such relations, forming a chained data structure that follows the dataflow. Specifically, it includes four subtypes:

\begin{itemize}[leftmargin=*]
    \item \textbf{\textit{Contains}} indicates that an entity $e_{to}$ is a member of another entity $e_{from}$, and can therefore be called from $e_{from}$ using a member access operator. Specifically, $e_{\text{from}}$ should be a \underline{\textit{Class}}, while $e_{\text{to}}$ can be one of the following types: a \underline{\textit{Function}}, an \underline{\textit{Attribute}} or a nested \underline{\textit{Class}}.

    \item \textbf{\textit{Returns}} indicates that an instance of a \underline{\textit{Class}} $e_{\text{to}}$ can be obtained by calling another entity $e_{\text{from}}$. Specifically, $e_{\text{from}}$ can be either a \underline{\textit{Function}} or an \underline{\textit{Attribute}}.
    
    \item \textbf{\textit{As Parameter}} indicates that an entity $e_{\text{to}}$ relies on an instance of a \underline{\textit{Class}} $e_{\text{from}}$ as a prerequisite for its call. Specifically, $e_{\text{to}}$ refers to a \underline{\textit{Function}}.
    
    \item \textbf{\textit{Inherits}} indicates that a \underline{\textit{Class}} $e_{\text{from}}$ is a subclass of another \underline{\textit{Class}} $e_{\text{to}}$, and therefore $e_{\text{from}}$ can call the members of $e_{\text{to}}$.
\end{itemize}

    \item \textbf{\textit{Calls}} relation denotes that within the implementation of a \underline{\textit{Function}} $e_{\text{from}}$, another entity $e_{\text{to}}$ is called. Specifically, $e_{\text{to}}$ can be a \underline{\textit{Function}}, an \underline{\textit{Attribute}}, or a \underline{\textit{Class}}.

    \item \textbf{\textit{Imports}} relation denotes that a \underline{\textit{Function}} $e_{\text{from}}$ can directly call another entity $e_{\text{to}}$ (excluding itself) within the current code context. This encompasses two scenarios: $e_{\text{to}}$ has been imported via an \texttt{import} statement and $e_{\text{to}}$ and $e_{\text{from}}$ reside within the same module scope. 
    Specifically, $e_{\text{to}}$ can be a \underline{\textit{Class}} or a \underline{\textit{Function}}.
\end{enumerate}

Formally, RSSG can be described as $G = \{(h, r, t) \mid h, t \in \mathcal{E},\ r \in \mathcal{R}_{ST} \cup \mathcal{R}_C \cup \mathcal{R}_I\}$, where $\mathcal{E}$ denotes the set of entities, representing the main code elements in the repository.
$\mathcal{R}_{ST}, \mathcal{R}_C, \mathcal{R}_I$ represent the sets of structural semantic relations corresponding to the three different views, \textit{i.e.}, \textbf{S}tructural and \textbf{T}ype Dependency, \textbf{C}alls, and \textbf{I}mports. In each triplet $(h, r, t)$, the head entity $h$ points to the tail entity $t$ via the relation $r$. 
The RSSG serves as a crucial foundation for our retrieval and call chain prediction.
Unless otherwise specified, all mentions of entities and relations in the following text refer to those defined in the RSSG.

\subsubsection{Call Chain}

Based on RSSG, we define a call chain as $\mathcal{C} = (e_1, r_1, e_2, \ldots, r_{n-1}, e_n)$, where $\forall \ 1 \leq i < n, (e_i, r_i, e_{i+1}) \in G, r_i \in \mathcal{R}_{ST}$. \textbf{Unlike conventional call chains, the call chains in our work focus on the static entity relationships within call statement(s) under the same scope.}
Intuitively, in source code, a single call often explicitly or implicitly involves multiple entities. For example, consider a call \texttt{c.f(d)}, where \texttt{c} is an instance of class \texttt{C}, \texttt{d} is an instance of class \texttt{D}, and the return type of \texttt{f} is \texttt{E}. This call involves four entities: \texttt{C}, \texttt{D}, \texttt{E}, and \texttt{f}, which can be further organized into two chains along the direction of dataflow: \texttt{C} $\rightarrow$ \texttt{f} $\rightarrow$ \texttt{E} and \texttt{D} $\rightarrow$ \texttt{f} $\rightarrow$ \texttt{E}. The relationships among these entities are encoded in the \textit{structural and Type dependency} relations.
If there are other calls in the context that are directly related to this call—such as a call used to obtain the instance \texttt{d} of \texttt{D}—the call chain can be further extended. In other words, a sequence of directly related calls can be jointly represented using a call chain.

\section{Approach}

\begin{figure*}[t]
    \centering
    \setlength{\abovecaptionskip}{0.1cm}
    \includegraphics[width=\linewidth]{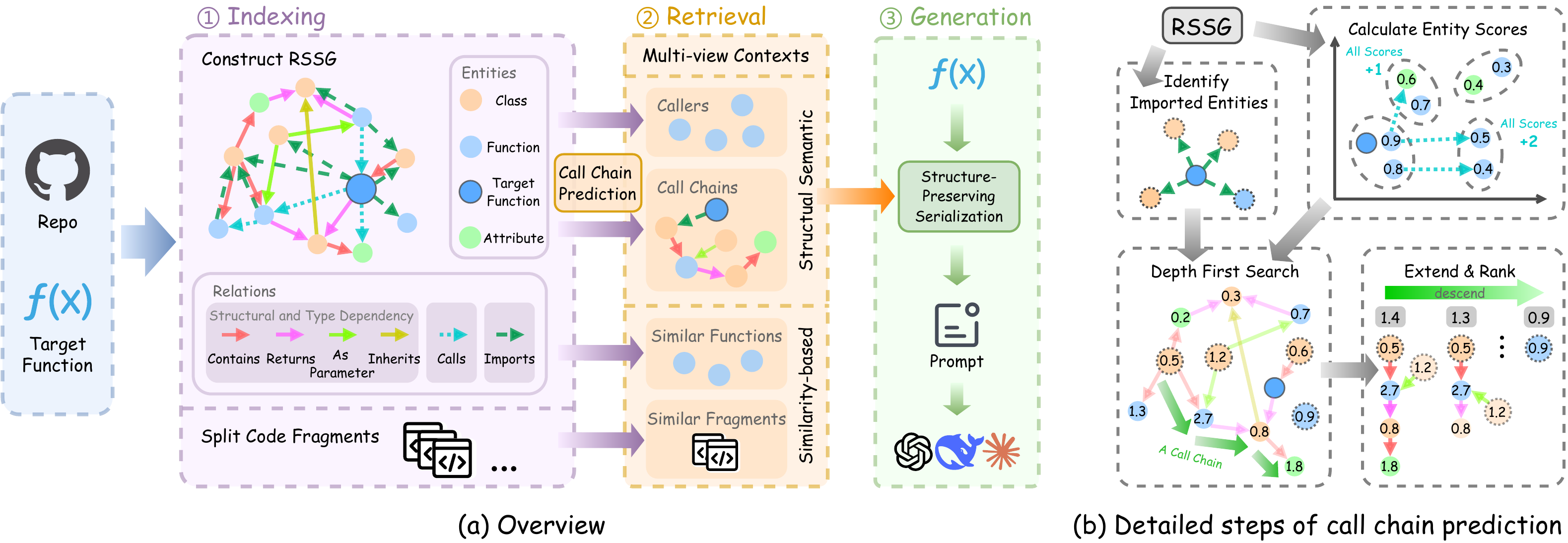}
    \caption{Workflow of \ourmodel{}.}
    \label{fig:overview}
    \vspace{-0.3cm}
\end{figure*}

\subsection{Overview}

With the observation in Section \ref{sec:motivating_example}, we proposed \ourmodel{}, a framework leveraging call chain-aware multi-view context for repository-level function generation. As illustrated in Figure \ref{fig:overview} (a), \ourmodel{} incorporates both the callers and the predicted callees of the target function into the LLM's prompt. It follows the standard RAG paradigm, which consists of three stages: \textit{indexing}, \textit{retrieval}, and \textit{generation}.
In the \textit{indexing} stage, we construct a unified RSSG that captures the semantic relationships among three types of code elements in the repository. This graph forms the foundation for retrieving both callers and callees. Additionally, we employ the sliding window algorithm \cite{zhang2023repocoder} to segment the repository code into fragments, which are used for similarity-based retrieval. In the \textit{retrieval} stage, we perform both retrieval and call chain prediction to obtain two types of contextual information from the index:  \textbf{structural semantic context} and \textbf{similarity-based context}. The former comprises \ding{172} function's callers and \ding{173} predicted call chain, while the latter includes \ding{174} functions and \ding{175} code fragments most similar to the unimplemented target function. Together, these form a four-view context representation. In the \textit{generation} stage, we design a serialization algorithm to reconstruct the call chain into a valid code structure while ensuring the elimination of duplicate nodes. Then we propose an iterative prompt construction strategy to systematically integrate the four context types into the prompt.
The final prompt is fed into the LLM to produce the function.

\subsection{RSSG Construction}
For each repository, we parse it using static code analysis tools to construct a RSSG. 
To support retrieval, we further process the RSSG accordingly.
Specifically, for each relation $r \in \mathcal{R}_C$, a weight $w_r$ is assigned to represent the number of times the head entity calls the tail entity. 
In addition, we compute an embedding vector $V_e$ for each entity $e$ using a sentence embedding model:

\vspace{-0.4cm}
\setlength{\jot}{2pt}
\begin{align*}
V_e& = \text{Embedder}( \\
        &<\text{name}>e_{\text{name}}</\text{name}><\text{signature}>e_{\text{sig}}</\text{signature}> \\
        &<\text{description}>e_{\text{docs}}</\text{description}><\text{path}>e_{\text{path}}</\text{path}> \\
    )   &
\end{align*}
where $e_{\text{name}}$, $e_{\text{sig}}$, $e_{\text{docs}}$, and $e_{\text{path}}$ represent the entity’s name, signature, docstring, and path, respectively. For example, for the target function in Figure \ref{fig:code_exp1}, these values are:

\begin{itemize}[leftmargin=*]
    \item $e_{\text{name}}$: \texttt{get\_deprecated\_args}
    \item $e_{\text{sig}}$: \texttt{def get\_deprecated\_args(self) -> collections.\\defaultdict}
    \item $e_{\text{docs}}$: \texttt{Returning dict with options which deprecate others}
    \item $e_{\text{path}}$: \texttt{infrared/core/inspector/inspector/SpecParser/\\get\_deprecated\_args}
\end{itemize}
If $e$ does not have a docstring, then $e_\text{docs}$ is set to an empty string.

\subsection{Call Chain Prediction}
\label{sec:call_chain_prediction}

Given the signature and docstring of an unimplemented target function $f$, we propose a heuristic method to predict the potential call chains that may appear in its function body based on RSSG. The core idea involves starting with the entities imported by the target function and performing a depth-first search on the RSSG to generate call chains from each search path. Each entity is then assigned a score based on the call patterns within the repository. These scores are aggregated to compute the score for each call chain, and the top-ranked call chains are selected as the final output. As illustrated in Figure \ref{fig:overview} (b), this method consists of four steps.

\noindent \textbf{(1) Identify Imported Entities.} 
First, we identify all entities (\textit{i.e.}, classes and functions) accessible directly from target function $f$, which can be directly extracted from RSSG:
$\mathcal{E}_{\text{imp}}(f) = \{ e \mid (f, r, e) \in G,\ r \in \mathcal{R}_I \}$.
These imported entities serve as initial starting points for call chain prediction. In the example from Figure \ref{fig:code_exp1}, the sole imported entity identified is the \texttt{SpecParser} class.

\noindent \textbf{(2) Calculate Entity Scores.}
Second, we compute entity scores using the call relation set $\mathcal{R}_C$ in the RSSG, where higher values indicate greater likelihood of being called. 
As mentioned in \ref{sec:motivating_example}, we could leverage the callees of similar functions to identify entities that are more likely to be called by the target function. 
To achieve this, we first apply the K-means clustering algorithm to partition all entities into clusters, {each containing approximately $M$ entities on average,} with the cluster containing entity $e$ denoted as $C_e$. We then compute a weighted score $S_e$ for each entity:

$$
S_e = \alpha_1 \cdot \text{sim}(V_f, V_e) + \alpha_2 \cdot \phi\left(\sum_{\begin{subarray}{c}(h, r, t) \in G,\ r \in \mathcal{R}_C,\ h \in C_f \setminus \{f\},\ t \in C_e \end{subarray}} \alpha_3 w_r\right)
$$

\noindent where $\text{sim}(\cdot)$ is vector similarity (cosine similarity), $\alpha_i (i =1,2,3)$ are weighting coefficients, and the function $\phi(\cdot)$ is a monotonically increasing concave function that mitigates sensitivity to large input.
For illustration, consider the top-right subfigure of Figure \ref{fig:overview} (b), where the number in each entity represents its similarity with the target function $f$. Ignoring $\alpha_i$ and $\phi(\cdot)$, each call of an entity $e$ by a function from $C_f$ increments the scores of all entities in $C_e$ by 1, added to their similarity with $f$.
Thus, an attribute entity with similarity 0.6 (as shown) would receive a final score of 1.6.

In summary, this algorithm computes entity scores by leveraging both repository call patterns and embedding similarity. Moreover, it enhances robustness by allowing entities within the same cluster to share call values (\textit{i.e.} the input of $\phi(\cdot)$). For example, in Figure \ref{fig:code_exp1}, if only \texttt{\_convert\_non\_cli\_args} or its callees were provided to the LLM, the model would struggle to infer that \texttt{iterate\_option\_specs} is a valid callee of the target function. 
However, because \texttt{iterate\_\\option\_specs} and \texttt{get\_option\_spec} reside in the same cluster, the former inherits the latter's call value, thereby boosting its retrieval probability.
It is important to note that, function embedding vectors ($V_e$) are computed without dependency on function bodies, which remain unavailable prior to code generation.

\noindent \textbf{(3) Depth First Search.}
Third, we perform a depth-first search (DFS) traversal on the $\mathcal{R}_{ST}$ relation set of RSSG starting from each entity in $\mathcal{E}_{\text{imp}}(f)$, collecting all possible call chains. By controlling the maximum chain length $l_{\text{max}}$, we can balance the breadth of retrieval with retrieval efficiency and precision. The algorithm pseudocode can be found in our replication package.

\noindent \textbf{(4) Extend \& Rank.} 
Finally, to enhance the LLM's understanding of each entity within the call chain, we extend the chain to include three key entities types (if exists):

\begin{itemize}[leftmargin=*]
    \item The constructor \underline{\textit{Function}} entity for each \underline{\textit{Class}} entity in the chain (using \textbf{\textit{Contains}} relation).
    \item The parameter and return type \underline{\textit{Class}} entities for each \underline{\textit{Function}} entity in the chain, as well as the type \underline{\textit{Class}} entity of each \underline{\textit{Attribute}} entity in the chain (using \textbf{\textit{As Parameter}} and \textbf{\textit{Returns}} relation).
    \item The \underline{\textit{Class}} entity to which each \underline{\textit{Function}} or \underline{\textit{Attribute}} entity in the chain belong (using \textbf{\textit{Contains}} relation).
\end{itemize}
We then compute each chain's overall score as the average score of its constituent entities, and the top $K_{\text{chain}}$ highest-scoring chains ($\mathcal{C}_{\text{chain}}$) are selected.
Since different call chains may contain each other—as seen in the first two call chains in Figure \ref{fig:overview} (b)—we deduplicate the chains during the ranking. To further enhance diversity, we select a maximum of $\tau$ chains starting from each entity in $\mathcal{E}_{\text{imp}}(f)$.

\subsection{Structure-Preserving Prompt Construction}
In addition to call chain prediction, we also retrieve three additional context types: (1) $K_{\text{caller}}$ caller entities of $f$ ($\mathcal{C}_{\text{caller}}$) with the minimum distances\footnote{If two entities reside in different files from $f$, the one with the shorter path length to $f$ in the file system tree is considered closer. Otherwise, the entity with a smaller line number difference from $f$ is considered closer.}; (2) the top $K_{\text{sim\_function}}$ function entities ($\mathcal{C}_{\text{sim\_function}}$) most similar\footnote{All similarity computations use the same embedding model and similarity function as those employed in call chain prediction.} to $f$ from RSSG; (3) the top $K_{\text{sim\_fragment}}$ code fragments ($\mathcal{C}_{\text{sim\_fragment}}$) most similar to $f$ from all code fragments. These contexts provide the LLM with relevant information for implementing the target function from both the perspectives of code similarity and structural semantics, forming a four-view context. Notably, $\mathcal{C}_{\text{sim\_fragment}}$ complements $\mathcal{C}_{\text{sim\_function}}$ by capturing similar code that may not reside within function definitions, thus preventing potential omissions.

Since entities within call chains may exhibit hierarchical relationships in the repository, we present a \textbf{structure-preserving serialization} algorithm to preserve their original structural organization during prompt construction, enabling the LLM to better capture these relationships and, as a result, improve its interpretation of the context. Additionally, because different call chains may contain overlapping entities, deduplication is necessary. Specifically, we traverse all the call chains to construct a structural tree, where each node's children are entities that share a \textbf{\textit{Contains}} relation with it. We then perform a preorder traversal to sequentially output all entities, using indentation to reconstruct their hierarchical structure. The detailed algorithm of the serialization strategy can be found in the replication package.

Considering the limited input token length of LLMs, we need to allocate the input budget reasonably across the four types of context. 
To achieve this, we propose a two-stage iterative prompt construction strategy.
Its basic idea aligns with the dynamic context allocation strategy \cite{shrivastava2023repository} but is more extensible. In short, each type of context is first allocated an equal token budget, after which each type ``claims'' any remaining tokens from the others according to a fixed priority order.
Specifically, given a maximum input length $\ell$ (after subtracting the length occupied by the task description and instructions) and $N$ types of context $C_1, \ldots, C_N$, our strategy proceeds as follows:

\noindent \textbf{Stage 1: Pre-allocation.} We first allocate a uniform budget of $\bar{\ell} = \left\lfloor \frac{\ell}{N} \right\rfloor$ to each context type $C_i$ {to balance the inclusion of different types of context. This averaged initialization strategy has been proven optimal in our preliminary experiments}. Then, $C_i$ fills $\bar{\ell}$ tokens with its context units according to its internal ranking strategy. If the shortest unit exceeds $\bar{\ell}$, the corresponding prompt is set to empty. Let $\ell_i$ denote the actual number of tokens used by $C_i$ after this step.

\noindent \textbf{Stage 2: Re-allocation.} Through preliminary experiments, we determined a fixed priority order over context types $C_{t_1}, \ldots, C_{t_N}$. For each context $C_{t_i}$ in order $i = 1, \ldots, N$, \ding{172} we allocate $\ell - \sum_{1 \leq j \leq N \land j \neq i} \ell_{t_j}$ tokens, allowing $C_{t_i}$ to claim remaining space of other contexts; \ding{173} similar to the first stage, $C_{t_i}$ refills the updated length with its context units; \ding{174} we then update $\ell_{t_i}$ accordingly.

Finally, the four types of context, together with the task description and generation instruction, are combined into a well-structured and properly formatted prompt, which is used to query the LLM for generating the target function. An example prompt is shown in Box \ref{box:prompt_example}. The 1st, 2nd, 3rd, and 4th contexts in the prompt correspond to the four contextual perspectives: similar code fragments, callers, call chains, and similar functions. Regarding call chains (3rd context), we can observe that entities retrieved via call chain prediction, such as \texttt{SpecDictHelper}, \texttt{get\_option\_spec}, and \texttt{iterate\_option\_specs}, are serialized in accordance with their structural organization in the repository code.

\begin{mybox}[box:prompt_example]{The prompt corresponding to the target function in Figure \ref{fig:code_exp1} (with omissions).}{t}
\textbf{You need to implement the function located at the end of the instruction based on relevant repository information.}

1. Here are some relevant code fragments from the repo:

\begin{lstlisting}[language=Python]
# infrared/common/library/virt_util.py
COMMANDS.setdefault(
    cmd_name, {'call': func, 'args': kwargs.keys()}
)
...

# infrared/__init__.py
...
\end{lstlisting}

2. Here are some functions in the repo that invoke the target function:

\begin{lstlisting}[language=Python]
# filepath: infrared/core/inspector/inspector.py, owning class: SpecParser
def validate_arg_deprecation(self, cli_args, answer_file_args):
    ...
    for deprecated, deprecates in self.get_deprecated_args().items():
        ...
...
\end{lstlisting}

3. Here are some relevant classes, functions, or attributes in the repo that you might use in the target function:

\begin{lstlisting}[language=Python]
# infrared/core/inspector/helper.py
class SpecDictHelper:
    """Controls the spec dicts and provides useful methods to get spec info."""
    ...
    def get_option_spec(self, command_name, argument_name) -> Any:
        """Gets the specification for the specified option name. """

    def iterate_option_specs(self) -> Generator[Tuple[dict, dict], Any, None]:
        ...
    ...
...
\end{lstlisting}

4. Here are some functions in the repo that are similar to the target function:

\begin{lstlisting}[language=Python]
# filepath: infrared/common/library/virt_util.py, owning class: Util
def _validate_args(self, *args):
    ...
    absent = []
    ...
...
\end{lstlisting}

\textbf{Please implement the following function:}

\begin{lstlisting}[language=Python]
from infrared.core.cli.cli import CliParser
from infrared.core.inspector import helper
...

# filepath: infrared/core/inspector/inspector.py, owning class: SpecParser
def get_deprecated_args(self):
	"""
	Returning dict with options which deprecate others.
	"""
\end{lstlisting}
\end{mybox}

\vspace{-0.12cm}
\section{Evaluation}

\subsection{Research Questions}

In this work, we aim to answer the following research questions:

\begin{itemize}[leftmargin=*]
    \item \textbf{RQ1: Overall Performance}. How does \ourmodel{} perform on the task of repository-level function generation?
    \item \textbf{RQ2: Ablation Study}. To what extent do the key components and strategies of \ourmodel{} contribute to its performance?
    \item \textbf{RQ3: Generalization Capability}. How generalizable is \ourmodel{} to other repository-level code generation tasks?
    \item \textbf{RQ4: Integration with Existing Approaches}. Can \ourmodel{} be integrated with existing repository-level code generation approaches to achieve further performance improvements?
\end{itemize}

\vspace{-0.26cm}
\subsection{Baselines}

We selected several representative repository-level code generation methods, including the current state-of-the-art, to conduct a performance comparison against \ourmodel{}.

\begin{itemize}[leftmargin=*]
    \item \textbf{Direct} refers to generating code solely based on the local context without incorporating any retrieval.
    \item \textbf{SimpleRAG} retrieves a fixed set of similar code snippets based on similarity and directly appends them to the prompt for one-shot generation.
    \item \textbf{RepoCoder} \cite{zhang2023repocoder} is a retrieval-augmented framework for repository-level code completion that combines a similarity-based retriever with a LLM in an iterative retrieval-generation pipeline. 
    \item \textbf{DRACO} \cite{cheng2024dataflow} is a repository-level code completion method that leverages extended dataflow analysis to construct a repo-specific context graph and enhances prompt by integrating structurally relevant background knowledge. 
    \item \textbf{CodeAgent} \cite{zhang2024codeagent} is an agent-based framework that equips LLMs with external programming tools to enable effective repository-level code generation. 
    \item \textbf{RLCoder} \cite{wang2024rlcoder} is a reinforcement learning–based framework for repository-level code completion that optimizes retriever behavior through perplexity feedback. 
\end{itemize}

For \ourmodel{} and each baseline, we evaluate their performance on four advanced backbone LLMs: GPT-4o mini\footnote{We use its latest model snapshot \texttt{gpt-4o-mini-2024-07-18}} \cite{hurst2024gpt}, Claude‑3.5‑\\Haiku\footnote{We use its latest model snapshot \texttt{claude-3-5-haiku-20241022}} \cite{anthropic2024claude}, Qwen3-235B-A22B\footnote{We use its ``non-thinking'' mode} \cite{yang2025qwen3} and DeepSeek-V3\footnote{We use its latest version \texttt{DeepSeek-V3-0324}} \cite{liu2024deepseek}.

\begin{table*}[t]
\centering
\small
\setlength{\abovecaptionskip}{0.1cm}
\caption{Pass@1 scores across datasets and backbone models, along with the average length of input tokens in each method. The percentages indicate the relative improvement compared to the best-performing baseline (underlined) under the same setting.}
\label{tab:rq1_result}
\begin{threeparttable}
\begin{tabular}{ll|llll|c}
\toprule
\multirow{2}{*}{\textbf{Dataset}} & \multirow{2}{*}{\textbf{Method}} & \multicolumn{4}{c|}{\textbf{Backbone Model}} & \multirow{2}{*}{\textbf{Avg. \# Token}} \\
& & \textit{GPT-4o mini} & \textit{Claude-3.5-Haiku} & \textit{Qwen3-235B-A22B} & \textit{DeepSeek-V3} & \\
\midrule
\multirow{6}{*}{\textit{CoderEval}} & Direct & 20.29 & 26.09 & 23.19 & 28.02 & 97 \\
& SimpleRAG & 39.13 & 43.96 & 43.00 & 48.79 & 3824 \\
& RepoCoder & 37.68 & 44.93 & \underline{44.44} & \underline{50.72} & 7661\tnote{1} \\
& DRACO & \underline{40.10} & 47.34 & 41.06 & 47.34 & 3837                      \\
& CodeAgent & 27.54 & 28.99 & 33.82 & 39.13 & 5884 \\
& RLCoder & 38.16 & \underline{49.28} & 40.58 & 46.86 & 3576 \\
\rowcolor{lightgray} & \textbf{\ourmodel{}} & \textsf{\fontseries{bx}\selectfont 44.93}~{\footnotesize\color{DeepGreen} $\uparrow$ 12.04\%} & \textsf{\fontseries{bx}\selectfont 55.07}~{\footnotesize\color{DeepGreen} $\uparrow$ 11.75\%} & \textsf{\fontseries{bx}\selectfont 49.28}~{\footnotesize\color{DeepGreen} $\uparrow$ 10.89\%} & \textsf{\fontseries{bx}\selectfont 59.42}~{\footnotesize\color{DeepGreen} $\uparrow$ 17.15\%} & 3679 \\
\midrule
\multirow{6}{*}{\textit{DevEval}} & Direct & 7.69 & 11.71 & 10.56 & 12.40 & 90 \\
& SimpleRAG & 22.61 & 28.42 & 21.81 & 28.53 & 3803 \\
& RepoCoder & 20.84 & 29.28 & 22.96 & 28.93 & 7613 \\
& DRACO & \underline{25.26} & \underline{30.48} & \underline{29.39} & \underline{32.09} & 3804 \\
& CodeAgent & 15.33 & 23.13 & 21.70 & 25.89 & 5649 \\
& RLCoder & 19.69 & 26.35 & 22.39 & 27.90 & 3668 \\
\rowcolor{lightgray} & \textbf{\ourmodel{}} & \textsf{\fontseries{bx}\selectfont 26.18}~{\footnotesize\color{DeepGreen} $\uparrow$ 3.64\%} & \textsf{\fontseries{bx}\selectfont 41.56}~{\footnotesize\color{DeepGreen} $\uparrow$ 36.35\%} & \textsf{\fontseries{bx}\selectfont 30.02}~{\footnotesize\color{DeepGreen} $\uparrow$ 2.14\%} & \textsf{\fontseries{bx}\selectfont 35.82}~{\footnotesize\color{DeepGreen} $\uparrow$ 11.62\%} & 3679 \\
\bottomrule
\end{tabular}
\begin{tablenotes}
    \item[1] The input token lengths for the RepoCoder and CodeAgent method are calculated as the sum of the input token lengths across all dialogue rounds.
\end{tablenotes}
\end{threeparttable}
\vspace{-0.3cm}
\end{table*}

\subsection{Datasets and Metrics}
We evaluate the effectiveness of \ourmodel{} in repo-level function generation tasks using two widely adopted benchmarks: CoderEval \cite{yu2024codereval} and DevEval \cite{li2024deveval}. In addition, to assess the generalizability of \ourmodel{}, we further evaluate its performance on repo-level API generation tasks using RepoEval \cite{zhang2023repocoder} in Section \ref{sec:api_level_expr}. The detailed information of the benchmarks are as follows:

\begin{itemize}[leftmargin=*]
    \item \textbf{CoderEval} \cite{yu2024codereval} consists of 460 real-world Python and Java tasks {(half for each language)} designed to evaluate code generation across six levels of context dependency. We use its Python subset. Due to incorrect \texttt{file\_path} entries in some samples, which prevent us from locating the corresponding RSSG entities of the target functions, we exclude those samples, resulting in a total of 207 samples.
    \item \textbf{DevEval} \cite{li2024deveval} is a developer-annotated benchmark designed to evaluate LLMs' coding abilities in real-world repositories, featuring 1,825 samples from 117 repos. Due to the presence of samples whose reference solutions fail to pass the provided test suites, as well as samples with incorrect lineno annotations, we exclude those instances, resulting in a final dataset of 1,742 samples. The specific details of the excluded samples for the above two datasets can be found in our replication package.
    \item \textbf{RepoEval} \cite{zhang2023repocoder} is a repository-level benchmark covering line-, API-, and function-level code completion tasks across real-world repositories. In Section \ref{sec:api_level_expr}, we use its API-level subset, which contains 1,600 samples.
\end{itemize}

For CoderEval and DevEval, since they provide test suites, we evaluate the correctness of the generated code using the \textbf{pass@k} metric \cite{chen2021evaluating}. For RepoEval, we employ \textbf{EM (Exact Match)} and \textbf{ES (Edit Similarity)} \cite{levenshtein1966binary} metrics to assess the similarity between the generated code and reference solution. {These benchmarks include repositories of various scales. Among the 158 repositories involved in CodeEval and DevEval, 17 repositories (10.76\%) contain more than 100,000 lines of code, which are typically classified as large-scale repositories. Tasks associated with these repositories account for 320 out of the total 1,949 tasks (16.42\%).}

\subsection{Implementation Details}

We construct the RSSG with the assistance of the \texttt{Pytype}\footnote{\url{https://github.com/google/pytype}} library and \texttt{Tree-sitter}\footnote{\url{https://tree-sitter.github.io/py-tree-sitter}}. Except for RLCoder, which uses a custom fine-tuned version of \texttt{UnixCoder}\footnote{\url{https://huggingface.co/microsoft/unixcoder-base}} \cite{guo2022unixcoder} (126M parameters) as its retriever (embedder), all other methods employ \texttt{bge-small-en-v1.5}\footnote{\url{https://huggingface.co/BAAI/bge-small-en-v1.5}} \cite{chen2024bge} (33.4M parameters) as the embedding model. {It is worth noting that preliminary experiments have shown that different advanced embedding models have no significant impact on the performance of our method.}
For call chain prediction, {the average cluster size $M$ is set to 5.} The weighting coefficients $\alpha_1, \alpha_2, \alpha_3$ of $S_e$ are set to $1.0$, $2.0$, and $2.0$, respectively. The call value mapping function $\phi(\cdot)$ is defined as $\phi(x) = \log_2(x + 1)$. The maximum call chain length $l_{\text{max}}$ is set to $5$, and the maximum number of call chains per starting point $\tau$ is set to $4$. The number of retrieved items for the four contexts—$K_{\text{chain}}, K_{\text{caller}}, K_{\text{sim\_function}}, K_{\text{sim\_fragment}}$—are all set to $5$. All the above parameters were determined through preliminary experiments.
The maximum prompt token length $\ell$ is set to $4096$.

All LLMs used in our experiments are accessed via their online APIs. Except for the temperature parameter, which is set to $0$, all other decoding parameters are kept at their default values. Temperature controls the randomness of LLM outputs, and setting it to $0$ reduces decoding to greedy decoding, resulting in deterministic outputs. We set $k = 1$ in the pass@k.
To minimize the impact of prompt token length—a extraneous variable—on the results and allow us to focus on the quality and accuracy of the prompt, we adjusted parameters in baselines during evaluation to make sure their average prompt token lengths were approximately at the same level. The RepoCoder \cite{zhang2023repocoder} and CodeAgent \cite{zhang2024codeagent} are exceptions, as their multi-turn dynamic interaction with the LLM makes it difficult to control the overall prompt length. {Nevertheless, we tried to align the final-turn (\textit{i.e.}, the final code generation turn) prompt length with that of other methods.}

\section{Results and Analysis}

\subsection{RQ1: Overall Performance}
To evaluate the effectiveness of \ourmodel{}, we compare it against baselines using four backbone models on the CoderEval and DevEval datasets. The pass@1 scores and the average input token length are presented in Table \ref{tab:rq1_result}.
As seen in the results, \ourmodel{} consistently achieves the highest pass@1 scores in all backbone models, with comparable or fewer input tokens, showing substantial improvements over the best-performing baselines, while maintaining high efficiency.
Notably, the most prominent gain is observed on DevEval with the Claude-3.5-Haiku backbone model, where \ourmodel{} outperforms the best baseline DRACO by 36.35\%. The second-best result is achieved on the CoderEval dataset with DeepSeek-V3, showing a 17.15\% improvement. {We validated the improvements statistically by comparing \ourmodel{} against the best baseline DRACO across all backbone LLMs and benchmarks. A two-tailed test yields $p = 2.11 \times 10^{-9} \ll 0.05$, confirming that \ourmodel{}'s gains are statistically significant.} These results demonstrate that \ourmodel{}, by offering more relevant, comprehensive, and well-structured contextual information, effectively enhances the performance of LLMs in repository-level code generation tasks.

Additionally, we observe that, in most cases, the best-performing baseline is DRACO, which retrieves context via dataflow analysis. This further underscores the importance of leveraging repository structural semantics for code generation. Interestingly, despite having the second-highest average input token length, the agent-based approach CodeAgent performs poorly, even falling behind SimpleRAG. This is primarily due to the large scale and logical complexity of the repositories used in the experiments, which makes it challenging for the LLM agent to effectively identify truly relevant content from the vast array of potential classes or modules using integrated tools. This suggests that, compared to static analysis or similarity-based retrieval, LLM-driven retrieval may still face considerable limitations when dealing with complex repository-level tasks.

\vspace{3mm}
\begin{custommdframed}
\textbf{\textit{Answer to RQ1:}} \ourmodel{} consistently outperforms all baseline methods across all backbone models with comparable or fewer input tokens, achieving a relative improvement of up to 36.35\% in pass@1 score. This superior performance validates the effectiveness and robustness of the contextual information provided by \ourmodel{}, highlighting its ability to enhance code generation tasks at the repository level.
\end{custommdframed}
\vspace{0mm}

\subsection{RQ2: Ablation Study}
To assess the effectiveness of each key component and strategy in \ourmodel{}, we conduct an ablation study on the CoderEval dataset using two best-performing backbone models overall (Claude-3.5-Haiku and DeepSeek-V3). 
Specifically, the ablation experiments are divided into three parts.
First, we evaluate the contribution of each of the four context views in the prompt to the overall performance. We individually remove (1) the \textit{callers (w/o Ca)}, (2) the \textit{call chains (w/o CC)}, (3) the \textit{similar functions (w/o SF)}, and (4) the \textit{similar code fragments (w/o SCF)}, and then assess the impact on code generation performance.
Second, we validate the necessity of three key strategies in call chain prediction: weighted entity scoring, depth-first search, and call chain extension. We construct three corresponding variants by removing each strategy: (1) \textit{scoring entities solely based on similarity (w/o WES)}, (2) \textit{using only directly imported entities (w/o DFS)}, and (3) \textit{disabling call chain extension (w/o CCE)}.
Finally, to evaluate the effectiveness of our structure-preserving serialization algorithm for prompt construction, we replace the algorithm with a baseline that \textit{simply exports each entity in the call chain sequentially without preserving structural relationships (w/o SS)}.

As shown in Table \ref{tab:rq2_result}, the results show that, except for the removal of the call chain extension strategy with Claude-3.5-Haiku, eliminating any key component or strategy leads to a decrease in pass@1 score, confirming their effectiveness. Among them, removing the similar function context (SF) when using the DeepSeek-V3 model results in the largest performance degradation, with a decrease of 11.38\%. However, when factoring in changes in prompt token length, the caller (Ca) and call chain (CC) contexts contribute the highest utility per token consumed among the four context types under our experimental settings, while similar code fragments (SCF) contribute the lowest.
Additionally, when the structure-preserving serialization algorithm is removed (w/o SS), the pass@1 score drops despite a slight increase in prompt length. This highlights the importance of this algorithm in both reducing token usage and improving generation quality.

Furthermore, to examine the quality of the predicted call chains, we compute the F1 scores for the predicted callees from our strategy and a baseline that predicts callees solely based on similarity, under the same average number of predicted callees.
The results show that our method (F1=0.603) outperforms the baseline (F1=0.4352) by 38.56\%, providing strong evidence for the effectiveness of our call chain prediction strategy. {Notably, accurately predicting the callees of a target function is inherently challenging, since similar functions can only provide guidance rather than deterministically dictate which calls should appear. While the current error rate may still seem high in absolute terms, it already represents a substantial improvement over baselines, thereby yielding stronger contextual relevance overall.}

\begin{table}[t]
\centering
\small
\setlength{\abovecaptionskip}{0.1cm}
\caption{Ablation study results on CoderEval.}
\label{tab:rq2_result}
\begin{tabular}{r|ll|l}
\toprule
\multirow{2}{*}{\textbf{Method}} & \multicolumn{2}{c|}{\textbf{Model}} & \multirow{2}{*}{\textbf{Avg. \# Token}} \\
& \textit{Claude-3.5-Haiku} & \textit{DeepSeek-V3} & \\
\midrule
\rowcolor{lightgray} {\textbf{\ourmodel{}}} & 55.07 & 59.42 & 3679 \\
\midrule
{w/o Ca} & 52.66~{\footnotesize\color{DeepRed} $\downarrow$ 4.38\%} & 55.56~{\footnotesize\color{DeepRed} $\downarrow$ 6.50\%} & 3281~{\footnotesize $\downarrow$ 10.82\%}  \\
{w/o CC} & 51.20~{\footnotesize\color{DeepRed} $\downarrow$ 7.03\%} & 55.07~{\footnotesize\color{DeepRed} $\downarrow$ 7.32\%} & 3245~{\footnotesize $\downarrow$ 11.80\%} \\
{w/o SF} & 52.17~{\footnotesize\color{DeepRed} $\downarrow$ 5.27\%} & 52.66~{\footnotesize\color{DeepRed} $\downarrow$ 11.38\%} & 2797~{\footnotesize $\downarrow$ 23.97\%} \\
{w/o SCF} & 51.20~{\footnotesize\color{DeepRed} $\downarrow$ 7.03\%} & 57.00~{\footnotesize\color{DeepRed} $\downarrow$ 4.07\%} & 2542~{\footnotesize $\downarrow$ 30.91\%} \\
\midrule
{w/o WES} & 53.14~{\footnotesize\color{DeepRed} $\downarrow$ 3.50\%} & 57.49~{\footnotesize\color{DeepRed} $\downarrow$ 3.25\%} & 3645~{\footnotesize $\downarrow$ 0.92\%} \\
{w/o DFS} & 52.66~{\footnotesize\color{DeepRed} $\downarrow$ 4.38\%} & 57.00~{\footnotesize\color{DeepRed} $\downarrow$ 4.07\%} & 3669~{\footnotesize $\downarrow$ 0.27\%} \\
{w/o CCE} & 55.07~{\footnotesize $\downarrow$ 0.00\%} & 57.00~{\footnotesize\color{DeepRed} $\downarrow$ 4.07\%} & 3665~{\footnotesize $\downarrow$ 0.38\%} \\
\midrule
{w/o SS} & 54.11~{\footnotesize\color{DeepRed} $\downarrow$ 1.74\%} & 56.04~{\footnotesize\color{DeepRed} $\downarrow$ 5.69\%} & 3728~{\footnotesize $\uparrow$ 1.33\%} \\
\bottomrule
\end{tabular}
\vspace{-0.3cm}
\end{table}

\vspace{3mm}
\begin{custommdframed}
\textbf{\textit{Answer to RQ2:}} Each proposed key component and strategy in \ourmodel{} positively contributes to the overall performance. Considering the variation in token length, callers and call chains are the most important among the four types of context, and the structure-preserving serialization contributes to both generation quality and token efficiency. Compared to the baseline, our call chain prediction strategy is able to retrieve more accurate callees. 
\end{custommdframed}
\vspace{0mm}

\subsection{RQ3: Generalization Capability}
\label{sec:api_level_expr}

Although \ourmodel{} is designed for repository-level function generation, it can be easily adapted to other repository-level generation tasks, such as API-level generation. In the case of API-level generation, given a code snippet $c$ from the repository, the model needs to generate a complete API call $d$ that directly follows $c$. To address this, we consider two cases for the location of $d$: (1) if $d$ resides within a function $f$, we can reuse the logic of function-level generation; (2) otherwise, we treat $c$ as a function entity $f$ and use it for retrieval.
Moreover, considering that the contextual code $c$ in API-level generation may already contain some API calls, we can leverage this information. As previously mentioned, programming experience suggests that similar functions tend to exhibit similar call patterns. Similarly, calls that follows similar call sequences are also often similar. Inspired by this, we extend our call chain prediction method to incorporate this sequential pattern in the repository. Specifically, We add a ``sequence value'' term into $\phi(\cdot)$, which represents the degree of matching between the sequence formed by entity $e$ and the entities involved in existing calls within $c$, and the sequence patterns observed in the repository.

We conduct experiments on the API-level subset of RepoEval using DeepSeek-V3 as the backbone, comparing \ourmodel{} with baseline methods. The results, shown in Table \ref{tab:rq3_result}, reveal that \ourmodel{} still demonstrates a notable improvement over the best-performing baseline in both EM and ES metrics, with fewer input tokens. In particular, for the Exact Match score, it achieves a relative gain of 7.33\%, indicating that \ourmodel{} also holds considerable potential for other repository-level code generation tasks beyond function-level generation.

\vspace{3mm}
\begin{custommdframed}
\textbf{\textit{Answer to RQ3:}} \ourmodel{} generalizes effectively to other repository-level code generation tasks, achieving a 7.33\% relative improvement in EM over the best baseline on the API-level generation task. This demonstrates its potential to extend beyond function-level generation and tackle a broader range of repository-level code generation challenges.
\end{custommdframed}
\vspace{0mm}

\begin{table}[t]
\centering
\small
\setlength{\abovecaptionskip}{0.1cm}
\caption{Generalization experiment results on RepoEval.}
\label{tab:rq3_result}
\begin{tabular}{l|ll|c}
\toprule
\textbf{Method} & \textbf{EM} & \textbf{ES} & \textbf{Avg. \# Token} \\
\midrule
Direct & 22.06 & 51.20 & 1505 \\
SimpleRAG & 34.25 & 61.02 & 3965 \\
RepoCoder & 32.94 & 59.60 & 7930 \\
DRACO & 33.19 & 64.36 & 4073 \\
CodeAgent & 36.13 & 60.86 & 14153 \\
RLCoder & \underline{37.50} & \underline{66.59} & 4081 \\
\rowcolor{lightgray} \textbf{\ourmodel{}} & \textsf{\fontseries{bx}\selectfont 40.25}~{\footnotesize\color{DeepGreen} $\uparrow$ 7.33\%} & \textsf{\fontseries{bx}\selectfont 67.74}~{\footnotesize\color{DeepGreen} $\uparrow$ 1.73\%} & 4055 \\
\bottomrule
\end{tabular}
\vspace{-0.3cm}
\end{table}

\begin{table}[b]
\centering
\small
\setlength{\abovecaptionskip}{0.1cm}
\caption{Results of integrating \ourmodel{} with RLCoder and RepoCoder on CoderEval.}
\label{tab:rq4_result}
\begin{tabular}{r|l|c}
\toprule
\textbf{Method} & \textbf{Pass@1} & \textbf{Avg. \# Token} \\
\midrule
\rowcolor{lightgray} \textbf{\ourmodel{}}* & 53.14 & 3962 \\
{\footnotesize \textit{+RLCoder}} & 54.11~{\footnotesize\color{DeepGreen} $\uparrow$ 1.83\%} & 3975 \\
{\footnotesize \textit{+RepoCoder}} & 56.04~{\footnotesize\color{DeepGreen} $\uparrow$ 5.46\%} & 7927 \\
\bottomrule
\end{tabular}
\end{table}

\subsection{RQ4: Integration with Existing Approaches}

Due to the low coupling between the contextual views in \ourmodel{}, it can be seamlessly integrated with other repository-level code generation methods to achieve better performance. To this end, we experimented with integrating \ourmodel{} with RLCoder \cite{wang2024rlcoder} and RepoCoder \cite{zhang2023repocoder}. Specifically, since both RLCoder and RepoCoder focus on similarity-based full-text retrieval, we replaced the similar code fragment context in \ourmodel{} with the code retrieved by these two methods.  
To construct more discriminative comparative experiments, we removed the similar function context from \ourmodel{} as a baseline and appropriately increased the number of similar code fragments. We denote this variant as \ourmodel{}*.
We conduct experiments on the CoderEval dataset using the DeepSeek-V3 backbone model, and the results are shown in Table \ref{tab:rq4_result}.
As observed, \ourmodel{} achieved consistent performance improvements when integrated with both two approaches.
Notably, the integration with RepoCoder resulted in a 5.46\% increase in pass@1 score. This result further highlights the broad applicability and enhancement potential of \ourmodel{}. Meanwhile, this indicates that \ourmodel{} still faces limitations in similarity-based context retrieval, which presents an opportunity for further optimization in future work.

\vspace{3mm}
\begin{custommdframed}
\textbf{\textit{Answer to RQ4:}} \ourmodel{} can be effectively integrated with existing repository-level code generation methods to achieve performance improvements. This demonstrates the broad applicability and enhancement potential of \ourmodel{}, offering promising directions for future optimization.
\end{custommdframed}
\vspace{0mm}

\section{Discussion}

\subsection{Case Study}

\begin{figure}[t]
    \centering
    \setlength{\abovecaptionskip}{-0.2cm}
    \includegraphics[width=\linewidth]{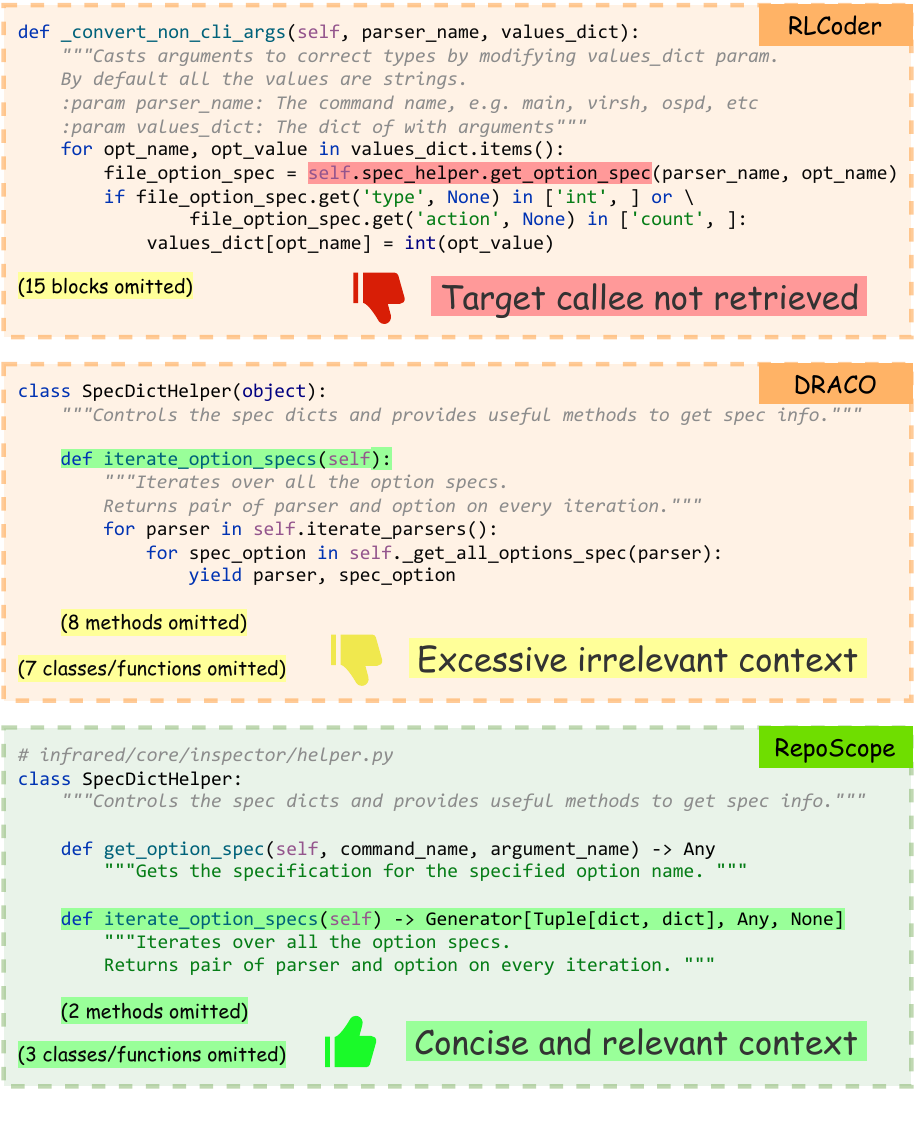}
    \caption{An example of the code snippets retrieved by \ourmodel{} and two baseline methods. The target function is the same as the one in Figure \ref{fig:code_exp1}.}
    \label{fig:case1}
    \vspace{-0.3cm}
\end{figure}

To intuitively illustrate the effectiveness of \ourmodel{}, we present an example in Figure \ref{fig:case1}, highlighting the differences in the retrieved context between \ourmodel{} and two baseline methods, RLCoder and DRACO. 
RLCoder, using a reinforcement-learned retriever, effectively retrieved a similar function \texttt{\_convert\_non\_cli\_args}. However, due to slight differences in the callees used by the target function and the retrieved one, \texttt{\_convert\_non\_cli\_args} fails to provide LLM with the complete set of relevant call information. DRACO, on the other hand, employs dataflow analysis and successfully locates the class \texttt{SpecDictHelper} and the method \texttt{iterate\_option\_specs}, both of which are called by the target function. However, it lacks a filtering mechanism for the retrieved code elements, leading to excessive irrelevant code being included in the context. For instance, the \texttt{SpecDictHelper} class alone introduces eight unused methods, which may obscure the truly relevant code and hinder the LLM's focus. 
In contrast, \ourmodel{} accurately predicts the callees of the target function by conducting in-depth static analysis and leveraging structural semantic information within the repository, while minimizing irrelevant context. This distinction largely explains why \ourmodel{} achieves superior performance in repository-level code generation tasks.

\subsection{Threats to Validity}
\label{sec:threats_to_validity}

\textbf{Threats to internal validity} arise from the hyperparameter settings used during the call chain prediction and the prompt construction process. We performed a small-range grid search over \ourmodel{}'s hyperparameters {on a small subset of data}, selecting the coefficients based on the best performance observed. It is anticipated that alternative hyperparameter configurations may lead to further improvements.
Additionally, due to budget constraints, our evaluation was restricted to a set of representative and advanced LLMs as the backbone models. Given the rapid progress in the field of LLMs, some models not included in this study may outperform those we evaluated. We will continue to monitor developments in this area and remain open to incorporating additional models in future work.
{Besides, the current use of pass@1 as the sole evaluation metric limits a comprehensive assessment of code quality, overlooking important attributes such as readability and maintainability. We plan to introduce LLM-as-Judge in the future to evaluate these attributes, providing a more comprehensive reflection of code quality.}

\noindent\textbf{Threats to external validity} relate to the generalizability of our approach and findings.  
Given the widespread adoption of Python language and the fact that most existing repository-level code generation benchmarks \cite{zhang2023repocoder, li2024deveval, yu2024codereval, ding2023crosscodeeval, li2024evocodebench, jimenez2023swe} are built using Python, we currently systematically evaluate our method only on this language. 
Nevertheless, compared to Python, statically typed languages (\textit{e.g.}, Java or C++) offer more explicit and easily extractable type relationships, which may make them even more amenable to our approach. {We conducted preliminary experiments on Java to validate this. Using the Java subset of CoderEval (229 tasks after excluding one faulty case) with GPT-4o-mini, \ourmodel{} outperformed all baselines that can run on Java despite only a basic adaptation (\textit{e.g.}, polymorphism not well-handled). Specifically, \ourmodel{} achieved a pass@1 of 58.95\%, whereas the best baseline, RLCoder, achieved only 57.64\%. We are willing to conduct a more systematic study of \ourmodel{}'s performance on such languages in future work.}

Moreover, our call chain prediction relies on the assumption that the call behavior of the target function can be inferred by similar functions within the same repository.
While this assumption holds in most cases, the prediction performance may degrade if the target function lacks similar counterparts (\textit{e.g.}, for small-scale repository){, if similar functions do not exhibit similar calling behaviors, or if the calling behaviors are sparse. Nevertheless, these issues generally do not occur in well-structured medium- to large-scale repositories. Additionally, even when call chain prediction is suboptimal,} contexts from other perspectives can still provide valuable semantic information to enhance code generation. {To support this, we conducted a validation on the 30\% most dissimilar tasks On CoderEval and DevEval benchmarks, determined by the similarity between each target function and its most similar counterpart in the repository. In these tasks across all four LLMs, \ourmodel{} still outperformed the best baseline DRACO with a 17.97\% relative gain, even higher than the 14.67\% improvement in all tasks, demonstrating robustness of \ourmodel{}.}

\section{Related Work}

Since the advent of programming languages, automated code generation has consistently attracted the attention of researchers \cite{olgaard2009automated,olgaard2010optimizations, barone2017parallel,becker2023programming,pujar2023automated,vadisetty2023leveraging}. With the rapid advancement of large language models (LLMs), many studies \cite{gee2024code,luo2023wizardcoder,huang2023enhancing,zheng2024opencodeinterpreter,zhong2024debug} have leveraged LLMs to generate code for programming exercises or competitive programming tasks, achieving remarkable performance. In recent years, repository-level code generation has garnered increasing attention due to its closer alignment with real-world software development scenarios. This growing interest has led to the introduction of various datasets, such as RepoEval \cite{zhang2023repocoder}, CoderEval \cite{yu2024codereval}, DevEval \cite{li2024deveval}, SWE-Bench \cite{jimenez2023swe}, \textit{etc.}, further advancing the field.

To address the challenge of incorporating extensive repository knowledge within the limited context window of LLMs, RAG techniques \cite{gao2023retrieval} have been introduced into repository-level code generation.
Early approaches \cite{zhang2023repocoder, shapkin2023dynamic} treated code repositories as natural language corpora, employing retrievers to identify a set of code snippets that are most similar to the generation context. These snippets then serve as relevant background knowledge for the LLM. Building on this foundation, subsequent research has focused primarily on three aspects: improving the accuracy of retrieving useful information \cite{liu2024graphcoder, deng2024r2c2, wang2024rlcoder, gao2024preference}, reducing unnecessary retrieval \cite{di2403repoformer, wang2024rlcoder}, and transforming the retrieved code snippets to make them more interpretable for the LLM \cite{shrivastava2023repofusion, gao2024preference}.

However, similarity-based retrieval methods do not always succeed in identifying the most relevant information. Due to the formal and structured nature of programming languages, code repositories contain rich program structural semantics which can serve as valuable signals for retrieval. 
To address this, some research has extended pure similarity-based retrieval. Among them, RepoFuse \cite{liang2024repofuse} enriches the LLM context by including the code of imported classes and functions. 
DRACO \cite{cheng2024dataflow} goes further by performing data flow analysis to identify local import information. 
Repohyper \cite{phan2024repohyper} constructs a Repo-level Semantic Graph and uses a Graph Neural Network (GNN) to compute node relevance scores for retrieval. Different from these approaches, CoCoGen \cite{bi2024iterative} adopts an iterative generation strategy, where it uses feedback from the interpreter to determine which code elements to retrieve. While these methods improve generation performance to varying degrees, they still fall short in fully leveraging program structural semantics. As a result, \ding{172} they may retrieve a large number of irrelevant codes \cite{liang2024repofuse, cheng2024dataflow}, or \ding{173} they have to rely on annotated data for training \cite{phan2024repohyper}, which can reduce performance and efficiency. 

Recently, agent-based approaches have gained increasing attention for their strong performance in repository-level code generation. CodeAgent \cite{zhang2024codeagent} integrates five programming tools and implements four agent strategies, enabling the LLM to determine retrieval targets autonomously.
LingmaAgent \cite{ma2024alibaba} constructs a knowledge graph for the repository and combines Monte Carlo Tree Search with LLM-based evaluation to identify the most relevant code snippets. These snippets are then integrated into the agent's pipeline. 
However, these approaches face notable limitations in terms of cost and efficiency due to the need for multiple LLM invocations.

Compared to existing approaches, \ourmodel{} leverages in-depth analysis of repository structural semantics to obtain more relevant and diverse contextual information. {Specifically, the core novelty of our approach lies in the call-chain prediction module. This module is the first to explicitly focus on potential callees of the target function as a distinct context type and achieves relatively accurate retrieval of such entities. More fundamentally, our method is the first to leverage the entire repository's structural information to enhance structure-based retrieval. Existing approaches only consider local information around the target function. For instance, DRACO restricts retrieval to code elements directly accessible at the completion site, overlooking non-local but potentially crucial elements (\textit{e.g.}, structurally similar functions). In contrast, our method expands the horizon: in the second step of call-chain prediction, we incorporate the call behavior of similar functions across the repository to compute entity scores. This design makes our retrieval both broader in scope and more robust than prior methods.} Moreover, \ourmodel{} relies solely on static analysis, {requires no training, and each function generation only invokes the LLM once after RSSG construction,} making it more advantageous in terms of cost and time efficiency. {This contrasts with training-based methods, which may need full retraining when components change, and agent-based methods, which repeatedly query LLMs at runtime, incurring higher cost and latency.}

\section{Conclusion and Future Work}

In this paper, we propose \ourmodel{}, a novel framework leveraging call chain-aware multi-view context for repository-level code generation. \ourmodel{} retrieves context from four distinct perspectives, providing LLMs with comprehensive repository background knowledge. Building upon our constructed Repository Structural Semantic Graph, we propose an effective call chain prediction method that enables the retrieval of more relevant contextual information. We further introduce a structure-preserving serialization algorithm to preserve hierarchical organization in prompts, making them more interpretable to LLMs. Our training-free, single-query approach is highly efficient. Extensive evaluation results demonstrate that \ourmodel{} substantially outperforms state-of-the-art baselines on repository-level code generation benchmarks, while exhibiting strong generalization and integration capabilities.

In addition to those mentioned in Section~\ref{sec:threats_to_validity}, we also plan to pursue the following directions in future work:  
\ding{172} Investigate how to infer specific user intentions from broader repository contexts when only vague requirements are provided.  
\ding{173} Explore alternative code snippet segmentation strategies (\textit{e.g.}, syntax-based segmentation) and study their impact. 
\ding{174} Investigate more fine-grained call-chain prediction techniques to further enhance accuracy.  
\ding{175} Explore deeper integrations with agent-based frameworks, such as enabling an agent to dynamically decide which contextual views to retrieve for a given target function, or exposing the call-chain prediction module as a callable tool for agents.
\begin{acks}
This research is supported by the National Natural Science Foundation of China Grants Nos. 62302021 and 62177003.
\end{acks}

\bibliographystyle{ACM-Reference-Format}
\bibliography{ref}


\end{document}